\documentstyle[aps,epsfig]{revtex}

\begin{document}
\draft

\title{\bf Study of Two-Step Mechanisms in Pion Absorption on $\bf ^6Li$, $\bf
^{12}C$ via Deuteron Emission} 

\author{G.M.~Huber\footnotemark, G.J.~Lolos, Z.~Papandreou, J.~Hovdebo,
S.I.H.~Naqvi} 
\address{Department of Physics, University of Regina, Regina, SK  S4S~0A2 
Canada} 

\author{D.F.~Ottewell, P.L.~Walden}
\address{TRIUMF, Vancouver, BC  V6T~2A3   Canada}

\author{G.~Jones}
\address{Department of Physics, University of British Columbia, Vancouver, BC  
V6T~1Z1   Canada} 

\author{X.~Aslanoglou\cite{greece}}
\address{Department of Physics and Astronomy, Ohio University, Athens, OH  
45701-2979} 

\date{\today}
\maketitle

\setcounter{footnote}{1}
\renewcommand{\thefootnote}{\fnsymbol{footnote}}
\footnotetext{Corresponding Author: TEL: 306-585-4240, FAX:
306-585-5659, E-mail: huberg@uregina.ca}

\begin{abstract}

The $(\pi^+,pd)$, and $(\pi^+,dd)$ reactions were investigated
with pions of 100 and 165 MeV kinetic energy on $^6Li$ and $^{12}C$ targets.
In comparison with previously published $(\pi^+,pp)$ data on the same targets
and at the same beam energies, kinematic regions were identified in which the
neutron pickup process $n+p\rightarrow d$ dominated the observed deuteron
yield.  The importance of this mechanism increases with energy, contributing
half of the observed cross section at 165 MeV.  The contribution of direct
quasi-triton absorption is significant only at 100 MeV.
\end{abstract}

\pacs{25.80.Ls,25.80.Hp,25.10.+s}

\section{Introduction}

Pion absorption reactions have been studied for many years because of their 
importance in understanding the pion-nucleus interaction in general.  Much of 
this effort has been spent in trying to understand the underlying pion 
absorption mechanism.  Because of energy and momentum conservation arguments, a 
minimum of two nucleons in the nucleus must be involved in the pion absorption 
process.  However, progress in understanding the answer to questions such as
the relative importance of absorption mechanisms involving three or more
nucleons, and the role of two-step mechanisms, has been difficult, with 
experiments often giving conflicting or incompatible signals.

With few exceptions, previous experiments have concentrated on the detection 
of energetic protons or neutrons emitted after pion absorption.  They have 
indicated that two-nucleon absorption is the dominant absorption mechanism at 
low energy ($T_{\pi} \approx$ 50 MeV) and that its importance drops with 
energy, accounting for less than 40\% of the absorption cross section by 
$T_{\pi}=240$ MeV \cite{ashery}.  Three-nucleon absorption increases in
importance with energy, and four-nucleon processes must be taken into account
for nuclei as heavy as $^{12}C$.  Recently, LADS has also found
conclusive evidence of the role of two-step mechanisms in the pion absorption
process \cite{lads}.

A small fraction of recent pion absorption experiments have also paid some
attention to the deuterons emitted in coincidence with a proton (i.e. 
($\pi^+,pd)$).  These experiments have been of interest
because direct absorption on a $pnn$ cluster can be cleanly identified without 
recourse to error-prone extrapolations over large regions of phase-space, and 
because the role of a second-step final state interaction (FSI) can be studied 
via the nucleon pickup process $p+n\rightarrow d$.  A synopsis of the most
representative works follows.

An early study on $^7Li$ and $^{12}C$ was published in 1978 by Comiso et
al. \cite{comiso}.  The effects of absorption on two nucleons were clearly
visible in the Dalitz plots formed from both elements, but effects due to
absorption on larger clusters could only be found in $Li$.  In 1986, Wharton et
al. \cite{wharton} investigated the $(\pi^+,pd)$ reaction on $^{6,7}Li$ at
$T_{\pi}=59.4$ MeV over a narrow angular range using a high resolution crystal
spectrometer.  Bound states of $^{3,4}He$ dominated their excitation spectra,
and they found that a model in which the pion interacts and is absorbed on a
``quasi-triton'' cluster described the main features of their data very
well. Yokota et al. published two studies \cite{yokota86,yokota89} obtained
with a low resolution scintillator array.  In the first, angular correlations
for the $(\pi^+,pp)$ and $(\pi^+,pd)$ reactions on $^{nat}C$ at $T_{\pi}=65$
MeV were compared, and it was found that the ratio
$d\sigma(\pi^+,pd)/d\sigma(\pi^+,pp)\approx 0.025$ was more constant with angle
than expected if the $pd$ were due to direct three-nucleon absorption.  In the
second, more detailed study on $^{6,7}Li$, it was found that at $T_{\pi}=165$
MeV, the $(\pi^+,pd)$ angular distribution is well reproduced by the
quasi-triton model at backward deuteron angles, $\theta_d>120^o$, but is higher
than the model by at least 300\% for $\theta_d<60^o$, indicating the
contribution of more complicated reaction mechanisms.

More recently, the BGO Ball Collaboration published ``detected'' cross sections
for events involving one or more deuterons in coincidence with one or more
protons \cite{ransome90,ransome92}.  Data were obtained on
$^6Li,\ ^{12}C,\ ^{27}Al,\ ^{58}Ni,\ ^{118}Sn,\ ^{238}U$ at $T_{\pi}=50,$ 100,
150, and 200 MeV. They found that the angular distributions and correlations
for proton and deuteron emission are similar to those of two-proton emission,
suggesting that most deuteron emission is probably due to neutron pickup by
protons.  Bauer et al. \cite{bauer} studied $ppp$, $pd$, and $ppd$
emission following absorption of 65 MeV $\pi^+$ on $^{16}O$.  They found
evidence for backward-peaked direct deuteron emission, confirming the finding
of Yokota et al. that deuteron emission has different signatures at forward and
backward angles.  Finally, the results of the LADS Collaboration are summarized
by Rowntree \cite{rowntree}.  Their experiment utilized a detector which
recorded coincidences between all types of outgoing particles, charged and
neutral, following $\pi^+$ absorption.  With this LADS detector, which
covered more than 98\% of $4\pi$, data were obtained for $^4He$, $N$, and $Ar$
gas targets, with a detection threshold of approximately 20 MeV for protons and
neutrons, and 25 MeV for deuterons.  They found that deuteron emission is
strongly correlated with the emission of other energetic charged (but not
neutral) particles, accounting for nearly 10\% of the total cross section on
$N$ when two charged particles are emitted, but approximately 50\% when four or
more charged particles are emitted.

In summary, previous results point to contributions from both quasi-triton
absorption and neutron pickup mechanisms.  While the total cross section of
pion absorption leading to deuteron emission is a small fraction of the total
absorption cross section, it is of particular interest because it can provide
additional information concerning the more conventional $pp$, $ppp$ and $ppn$
channels.  In the kinematic regions where $pd$ emission exhibits signatures of
quasi-triton ($pnn$) absorption, it can provide information on T=1/2 isospin
cluster composition in the ground state of nuclear systems.  The two-body
nature of the $pd$ final state may also make the extraction of such information
more certain than the three-body $ppn$ channel, whose extraction is based on
three-body phase-space extrapolations.

On the other hand, the $pd$ and $dd$ exit channels may provide better
information on the role of hard FSI (in terms of pick-up processes) than can be
provided by re-scattering or other less understood types of FSI.  Even though
FSIs have long been recognized as one of the complications in pion
absorption which can mimic multi-nucleon absorption mechanisms, such
interactions have been very difficult to identify and quantify with any
certainty and confidence. This problem has been partly due to the difficulty of
extracting FSI signatures from the many various channels available to them,
partly due to the broadening of the distributions due to the Fermi momenta of
the particles involved, and partly due to the multi-body phase-space
distributions that need to be unfolded and which are not unambiguous.  Thus,
although initial state interactions (ISI) in pion absorption have been
identified and quantified (Ref. \cite{lads,lehmann}), the clean identification
of FSI remains elusive.  A clear signature of FSI in terms of pick-up
reactions, both qualitatively and quantitatively, will greatly assist our
understanding of two-step FSI contributions in this and other
hadron final-state reactions.

In this work, we present results for the $(\pi^+,pd)$ and $(\pi^+,dd)$
reactions on $^6Li$ and $^{12}C$ targets at 100 and 165 MeV incident pion
energies.  This is the final of a series of papers \cite{pap95,lolos96,huber97}
resulting from a pion absorption experiment at TRIUMF.

\section{Experiment and Data Analysis}

$pp$ coincidence results following $\pi^+$ absorption on $^6Li$ and $^{12}C$
with this experimental apparatus have already been published
\cite{pap95,lolos96,huber97}.  These publications, as well as reference
\cite{pap88}, describe the apparatus and data analysis in considerable detail. 
Therefore, the experiment will only be briefly described here, with special
attention paid to those aspects of the analysis which are specific to the
detection of deuterons. 

The experiment was performed at the M11 pion channel at TRIUMF, using $\pi^+$ 
beams with kinetic energies of 100 and 165 MeV at the center of the target.
Pions were identified by time-of-flight measurements, and the total incident 
flux was corrected for the known $\mu^+$ and $e^+$ contaminations.  The
flux measurement was determined using a plastic scintillator counter positioned
upstream of the target which was normalized to another in-beam monitor having the 
same size as the target and placed directly in front of it, but removed during 
the data taking.  Typical beam rates were 3~MHz with pile-up corrections of 
approximately 8.5\% applied.

\begin{figure}[h]
\begin{center}
\epsfig{file=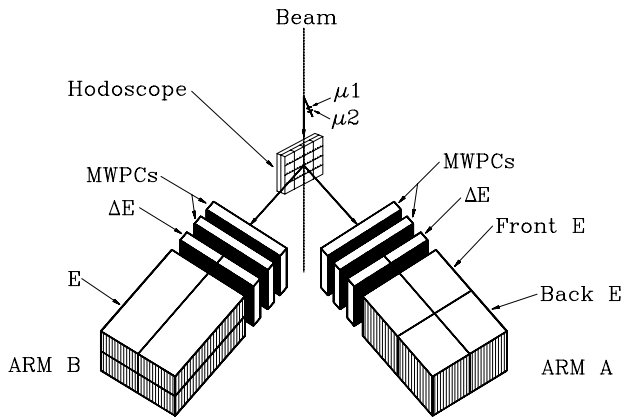,height=3.2in,angle=90}
\caption{Schematic diagram of the experiment.  Full details of the 
apparatus are discussed in reference \protect\cite{pap88}.}
\end{center}
\label{fig1}
\end{figure}

The targets, metallic $^6Li$ with a 95\% enrichment factor, and $^{nat}C$
(research grade graphite), were alternately mounted on the downstream side of a
segmented plastic scintillator hodoscope which measured the beam profile and
acted as an extra flux monitor.  As a result of the carbon content in the
$(CH_{1.1})^k$ scintillator hodoscope, a carbon subtraction had to be employed
to obtain the $^6Li$ data.  The hydrogen content of the hodoscope, however,
clearly did not contribute to the $(\pi^+,pd)$ reaction.  The areal densities
of the targets were 112 $\rm mg/cm^2$ for the $^6Li$ target, and 399 $\rm
mg/cm^2$ for the $^{nat}C$ target.  The carbon content of the hodoscope
contributed an additional 190 $\rm mg/cm^2$.

The scattered particles were detected in two detector arms composed of plastic
scintillators and multiwire proportional chambers (MWPC), shown
schematically in figure 1.  The two arms were located on opposite sides of the
pion beamline, and could be rotated independently around the center of the
scattering table between angles of $20^o$ and $150^o$ with respect to the
incident $\pi^+$ beam, measured from the physical edges of the detectors.  Each
arm was segmented into two subassemblies (Left and Right) with respect to the
incoming pion beam, with two common MWPCs in each arm, but with separate
$\Delta E-E$ detectors which provided the particle identification and energy
information.  Arm A was equipped with four 15 cm thick blocks, each viewed by
two photomultiplier tubes (PMTs) via $45^o$ total reflection prism light
guides, while Arm B consisted of four 30 cm thick monoblocks, each viewed by a
single PMT.  The $\Delta E$ counters were viewed by two PMTs each.  This
configuration resulted in a 220 MeV maximum stopping power of the detectors for
protons \cite{pap88}, which translates to a range of 30 cm.  The MWPCs provided
particle trajectory information, allowing a three-dimensional traceback of each
event to the target.

The trigger was defined by the four-fold coincidence $\Delta E_A \cdot E_A
\cdot \Delta E_B \cdot E_B$, in any of the four possible combinations of the
subassemblies, as long as one charged particle was recorded on each side of the
beam.  For Arm A, $E_A$ was defined by the front $E$ counter only, with the back 
$E$ counter not included in the trigger decision.  Thus, for a pair of central
angles of each arm, data were collected simultaneously for four effectively
independent telescope coincidence combinations.  As the angular acceptance of
each arm was $\Delta\theta=22^o$, each coincidence combination involved $11^o$
in each arm.  Data were obtained for a variety of subassembly angle
combinations between $22^o$ and $145^o$ on each side of the beam.  The angular
resolution of the reconstructed event was $1.6^o$, taking into account the MWPC
tracking resolution of each arm and the divergence of the incident pion beam.
The light output response and normalization of the detectors and electronics
were calibrated and periodically monitored throughout the run via the known
proton energies from the $d(\pi^+,pp)$ reaction at several proton conjugate
angles at 100 and 165 MeV incident pion energies, as described in
\cite{pap95,lolos96,huber97}.

\begin{figure}[h]
\begin{center}
\epsfig{file=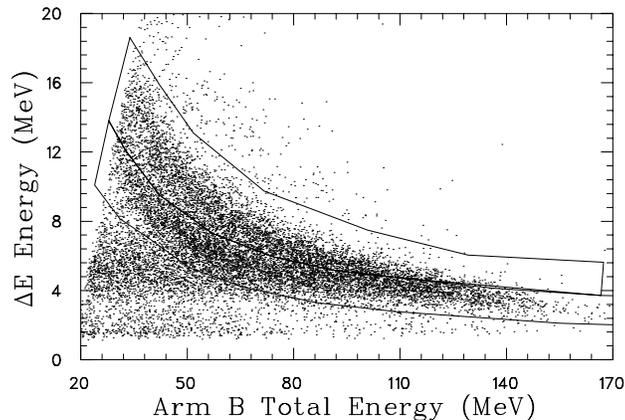,height=3.2in,angle=90}
\caption{A representative particle identification plot from the experiment.  
Bands due to proton and deuteron emission are clearly visible.}
\end{center}
\label{fig2}
\end{figure}

A representative particle identification (PID) plot is shown in figure 2,
showing that pions and/or muons, protons, and deuterons were clearly separated.
In Arm A, energetic protons penetrating the front counters and stopping in the
rear counters formed additional PID bands, providing a greatly enhanced
separation of pions and/or muons from the protons.  In addition to the periodic 
$d(\pi^+,pp)$ runs, the gain stability of each detector was verified for the
data obtained at each angle pair setting by checking to make sure that the
proton band remained at its calibrated position.  The offline software gain 
parameters were occasionally adjusted as needed.  

Particles located within the upper and lower boxes of figure 2 were identified
as deuterons, and protons, respectively.  Events which fall far to the left of
the proton band correspond to particles which either suffered nuclear
reactions in the detector (``reaction losses'') or did not stop in the
detector due to the initial trajectory or multiple scattering (``edge
effects'').  Energy-dependent corrections were made for these effects, based on 
the ratio of events to the left of the proton band to the number of events in 
the proton band.  This correction is described in more detail in reference
\cite{huber97}.  Typical factors, for the energy region of highest population
(135 to 165 MeV), are 1.20 to 1.35, respectively.  No reaction loss correction
was applied to deuterons as there was no independent method of determining its
value.

It has been shown that the light output versus energy deposition depends on
whether the particles were traversing or stopping in the scintillators
\cite{pap88a}.  For arm A, where the thick front blocks could either stop
protons or allow them to be transmitted to the rear blocks, such differences in
light output were readily observed and easily corrected for. These corrections
were compared to the well-defined proton energies expected from the
$d(\pi^+,pp)$ reaction, and were found to be in excellent agreement. An
additional correction was applied to the light output from the stopping counter
if the particle was identified as a deuteron, in order to adjust for
the difference in the light output of the deuteron relative to the proton
because of the different ionization density and subsequent quenching in the
scintillators.  The light output curve was parameterized by a quadratic
polynomial and an exponential equation according to the light output data
provided in reference \cite{gooding}.

Instrumental energy resolutions of 2.9\% and 2.6\% $(\sigma)$ at incident pion
energies of 100 and 165 MeV, respectively, were achieved, once all corrections
were made.  These numbers translate to approximately 4 MeV $(\sigma)$ for the
resolution in the excitation spectra.  A uniform detection threshold of 45 MeV
for protons, and 50 MeV for deuterons was imposed to avoid threshold variations
resulting from the different target angle orientations and the associated
variation in the energy loss in the target.  MWPC efficiencies, determined on a
run-by-run basis and in a cyclic manner for each MWPC, were also applied to the
data.

Finally, the geometrical solid angle of each two-arm coincidence-pair
was corrected for the effects of multiple scattering, finite beam spot size,
and limited vertical acceptance, by implementing an acceptance correction
factor
\begin{displaymath}
d\Omega_1 d\Omega_{2\ expt}=d\Omega_1 d\Omega_{2\ point}*Acc ,
\end{displaymath}
where $d\Omega_1 d\Omega_{2\ point}$ are the point source solid angles
of the front face of the $E$ counters in each arm subassembly consistent with
the data analysis.  The factor $Acc$ (equal to $0.80\pm 0.07$ at both incident
pion energies) was determined using a Monte Carlo in which three-energetic
emerging body phase space is assumed for the out-of-plane distribution.  After
application of this factor, the experimental data may be reliably integrated
over $4\pi$.

\section{Monte Carlo models}

\subsection{Motivation}

\begin{figure}[h]
\begin{center}
\epsfig{file=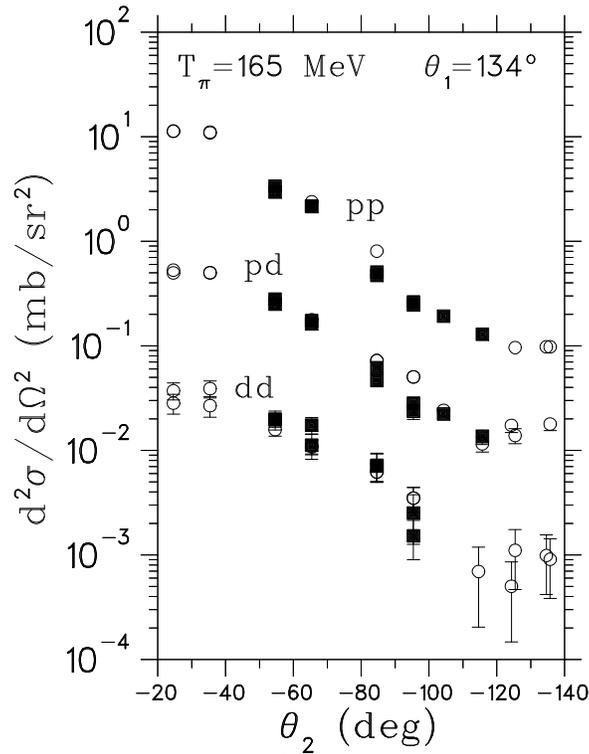,width=3.0in}
\caption{The open circles are $^{12}C$ and the filled squares are $^6Li$
target, respectively.  It should be emphasized that the angular distributions
are not arbitrarily offset from each other, but reflect the actual cross
sections for each process.}
\end{center}
\label{fig3}
\end{figure}

Representative angular distributions of the data are shown in figure 3.  Two
striking features emerge.  The first is that the angular distributions for the
$pp$, $pd$ and $dd$ final states are nearly identical in shape, with the $pd$
distribution slightly flatter than the $pp$ distribution, and the $dd$
distribution the flattest of all.  This similarity was reported earlier for
$pd$ emission from a variety of nuclei by Ransome et al. \cite{ransome92}
who concluded ``that most deuteron emission is probably due to neutron pick-up by
protons''.  Unfortunately, however, the poor angular resolution of the BGO Ball
data did not allow this hypothesis to be tested in a more quantitative manner.
The even flatter angular distribution characteristic of $dd$ emission is a new
observation, and serves to reinforce the argument for the prevalence of neutron
pickup.  In the case of figure 3, the decrease in cross-section as $\theta_2$
varies from $-25^o$ to $-145^o$ is a factor of 150 for $pp$, 42 for $pd$, and
only 16.5 for $dd$.

The second significant feature of figure 3 is that the differential cross section is
reduced by nearly the same factor of ten between $pp$ and $pd$ as
between $pd$ and $dd$, regardless of whether the emission is from $^6Li$ or
$^{12}C$.  This is especially noteworthy, because the $dd$ reaction on $^6Li$
leaves no remaining neutrons, while the reaction on $^{12}C$ leaves three
unobserved neutrons, and so the two reactions must have significantly different
neutron pickup combinatorics.

The extent to which neutron pickup could account for the nearly constant factor
between the different angular distributions can be obtained from a rough
estimate based on the total cross section (approximately 15 mb
\cite{pd30,pd52,pd65}) for the $^{12}C(p,d)^{11}C_{g.s.}$ process between
$T_p=30$ and 65 MeV.  If we naively assume that the $\pi^+$ absorption occurs
in the center of the $^{12}C$ nucleus, we immediately obtain a $p+n\rightarrow
d$ pickup probability of 4.6\%.  So, it is conceivable that a full calculation
could account for a large portion of the 10\% deuteron emission probability
observed in figure 3.  To study this in more detail, two Monte Carlo
simulations were employed.  These are described below.

\subsection{$\bf p+n\rightarrow d$ pickup}

This model simulates the production of a deuteron by pion absorption on
$^{12}C$ under the assumption that the absorption initially leads to the
emission of nucleons: two protons, which are tracked in the simulation, and
any remainder which are handled through appropriate setting of the missing
momentum and mass variables.  One of the protons combines with a neutron to
form the deuteron observed in the final state, while the other proton continues
to the detectors.  The effect of the various absorption mechanisms at the
initial absorption vertex, such as the absorption on two, three or four
nucleons, or soft initial- or final-state interactions, are taken into account
in an empirical manner via the method discussed below.

The CERN program, GENBOD \cite{genbod}, was used to create the momentum vectors
of the participating particles, two protons, one neutron, and recoiling
excited $^9B$, randomly chosen within the available phase-space of the
reaction.  To account for the various possible absorption mechanisms leading
to extra unobserved nucleons, a triple-differential cross section for the
events was employed, parameterized from the 100 and 165 MeV $^{12}C(\pi^+,pp)$
data of reference \cite{huber97}.  To simplify matters, the differential cross
section was assumed to be separable into functions over angle and excitation energy
\begin{equation}
\frac{d^3\sigma} {d\Omega_{p_1} d\Omega_{p_2} dE_{exc}} 
( T_{\pi}, \vec{p_{p_1}}, \vec{p_{p_2}} ) =
\frac{d^2\sigma}{d\Omega_{p_1} d\Omega_{p_2}}
( T_{\pi}, \theta{p_1}, \theta{p_2} ) .
\frac{dS} {dE_{exc}}
( T_{\pi},p_{p_1},p_{p_2},\Delta\theta_{QFA} ) ,
\end{equation}
where $\frac{dS}{dE_{exc}}$ refers to the normalized excitation energy spectra
for $pp$ emission, and $\Delta\theta_{QFA}$ is the difference between the
second detected particle angle and the angle expected from simple $d(\pi^+,p)p$
kinematics.  All other variables are standard, and should be self-explanatory.

For several values of the first proton angle, the $^{12}C(\pi^+,pp)$ angular
distributions, $\frac{d^2\sigma}{d\Omega_{p_1}d\Omega_{p_2}}$, were fit over
the second proton angle with the sum of two Gaussians, one narrow and one
broad, both having the same centroid.  The individual coefficients of
this fit were then parameterized in terms of Legendre polynomial functions of
the first proton angle.  The cross section for a particular event could then be
calculated, at pion energies of 100 MeV or 165 MeV, in terms of the scattering
angles of the two protons.  The choice of function for these fits was based
purely on achieving a good fit with few variables, and is not meant to imply
any additional physical significance.

The parameterization of $\frac{dS}{dE_{exc}}$ was based on the normalized
excitation energy spectra for proton emission from reference \cite{huber97}.
The shape of the excitation energy spectra is a function of the angle
difference between the observed second proton angle and the angle expected from
simple $d(\pi^+,p)p$ kinematics, $\Delta\theta_{QFA}$.  At zero difference
between the angles, there is a prominent peak for low excitation, due to
the large contribution from direct absorption on two protons, with the spectrum
falling off rapidly thereafter. As the difference increases further, a broader
distribution due to contributions from processes involving more nucleons
becomes part of the spectrum.  This addition causes a smearing of the
two-proton absorption peak, giving the distribution an overall Gaussian shape
with center at higher excitation energies.  The spectra were thus fit using a
piecewise function consisting of a Gaussian definition that gives way to a
linear term.  By varying the parameters of this function, including the point
where the definition of the function switches from Gaussian to linear, the
spectra could be fit quite well with the same function for all values of the
angular difference.

The simulation then assumed that one of the protons picked up a neutron, via a
hard final-state interaction, to form the deuteron observed in the final state.
The pickup of the neutron by a proton in the recoil nucleus was approximated by
the experimentally measured differential cross sections for the reaction
$^{12}C(p,d)^{11}C_{g.s.}$ as a function of $T_p, \theta_d$
\cite{pd30,pd52,pd65,pd100,pd156,pd185}, after appropriate Lorentz
transformation to the c.m. frame of the struck neutron, $^9B^*$ system.  This
nucleus was chosen, rather than $^{10}B$, because of the availability of
experimental data.  Comparison with the limited $^{11}B$ data available
indicate that this approximation should affect the model in any significant
manner.  As the cross section for the $^{11}C_{g.s.}$ final state is an order
of magnitude higher than for any of the other possible discrete final states,
it was sufficient to use just the ground state in the Monte Carlo.  The angular
distributions of differential cross sections $\frac{d\sigma}{d\Omega_d}$ were
fit with a sum of two Gaussians and a linear term.  The fit coefficients were
then obtained as a function of incident proton energy.  Since the
parameterization was based on $(p,d)$ cross sections from $T_p=30$ to 185 MeV,
it did not include the `soft' $^1S_0$ FSI, which dominates for relative $NN$
momenta less than 80 MeV/c.

Monte Carlo data were generated with statistical weight
\begin{displaymath}
w = \sqrt{ 
\frac{d^3\sigma}{d\Omega_{p_1} d\Omega_{p_2} dE_{exc}} 
( T_{\pi}, \vec{p_{p_1}}, \vec{p_{p_2}} )
\frac{d^2\sigma}{d\Omega_{d}dE_{p_2}}(\vec{p_{p_2}},\vec{p_n},\vec{p_d})
}
\end{displaymath}
and were then tracked through the target and detector assemblies, taking into
account the effect of detector acceptance and resolution.  Finally, the
simulated data were analyzed in the same manner as the experimental data.

$dd$ emission following $\pi^+$ absorption was similarly modeled via two
pickup reactions.  In this case, the initial input to GENBOD was modified so
that five particles were produced, two protons, two neutrons and a residual
$^8B$ nucleus.  The simulation then proceeded as before, but with each proton 
picking up a neutron to form the deuterons observed in the final state.

\subsection{``$\bf pp$-like'' $\bf pd$ emission}

What we term $pp$-like $pd$ emission was simulated in the following manner.
GENBOD was used to create the momentum vectors of three outgoing particles, $p$,
$d$, and $^9B^*$.  It was assumed that the triple differential cross section for
this process is given by the same parameterization as equation (1) above, after
accounting for modification due to the different phase-space characterizing the
$pd ^9B^*$ final state, instead of $pp ^{10}B^*$.  This simulation differs from
the quasi-triton $\pi^+t \rightarrow dp$ process because it incorporates the
contribution of more complex reaction mechanisms present in the
$pp$ emission data.  Thus, the $d$ could either originate directly from the
absorption vertex, or via knockout before absorption (ISI), or via an outgoing
nucleon soft FSI which does not materially affect the outgoing momentum
distributions.  

Although the two mechanisms B and C, overlap somewhat kinematically,
their respective energy and angle signatures are distinctive enough to allow
their relative contributions to be distinguished.  The comparison of the pickup
and $pp$-like models, therefore, provides a measure of the hard neutron pickup
contribution, where the momentum of the deuteron differs substantially from
that of the originally emitted nucleon.  Because these models are based upon
empirical data, it is only possible for us to compare their relative cross
sections, and not their amplitudes.  Therefore, interferences between the two
mechanisms are not taken into account in our analysis.  Since spin averaging 
and other similar effects are likely to wash out such amplitude interference 
effects, an argument can be made that this necessary approximation employed
will not greatly affect the conclusions of the analysis.

\section{Features of the data}

Using these two models, it was possible to obtain a remarkably good description
of all observables extracted from the data at 100 and 165 MeV pion energies
incident upon $^{12}C$.  Because each model incorporated empirical fits to
$^{12}C(\pi^+,pp)$ data, neither model was applied to the $^6Li$ target data.
There were a number of reasons for this restriction.  Firstly, a target
subtraction was required to extract the $^6Li$ data (section II), so it is of
poorer quality than the $^{12}C$ data.  Secondly, the significant contributions
of both the discrete and continuum final states make the excitation spectra
for $^6Li$ more difficult to parameterize than the smooth distributions
obtained with a $^{12}C$ target.  Finally, because either one or zero neutrons
are left unaccounted for after the reaction on a $^6Li$ target, the neutron momentum
distributions may not be independent of each other, resulting in additional
complications for the pickup model.  These considerations led to an easier
application of the pickup mechanism model to $^{12}C$ than to $^6Li$.
Nonetheless, the similarity of the results for the two nuclei, as shown in
figure 3, lead us to believe that the conclusions drawn from the $^{12}C$ data
will be largely applicable to the $^6Li$ data as well.  While each model is
based on a number of assumptions and simplifications, the conclusions are based
primarily on kinematic distributions which tend to be relatively insensitive to
details of the reaction mechanism.

\subsection{$T_{\pi}=165$ MeV data}

\begin{figure}[h]
\begin{center}
\epsfig{file=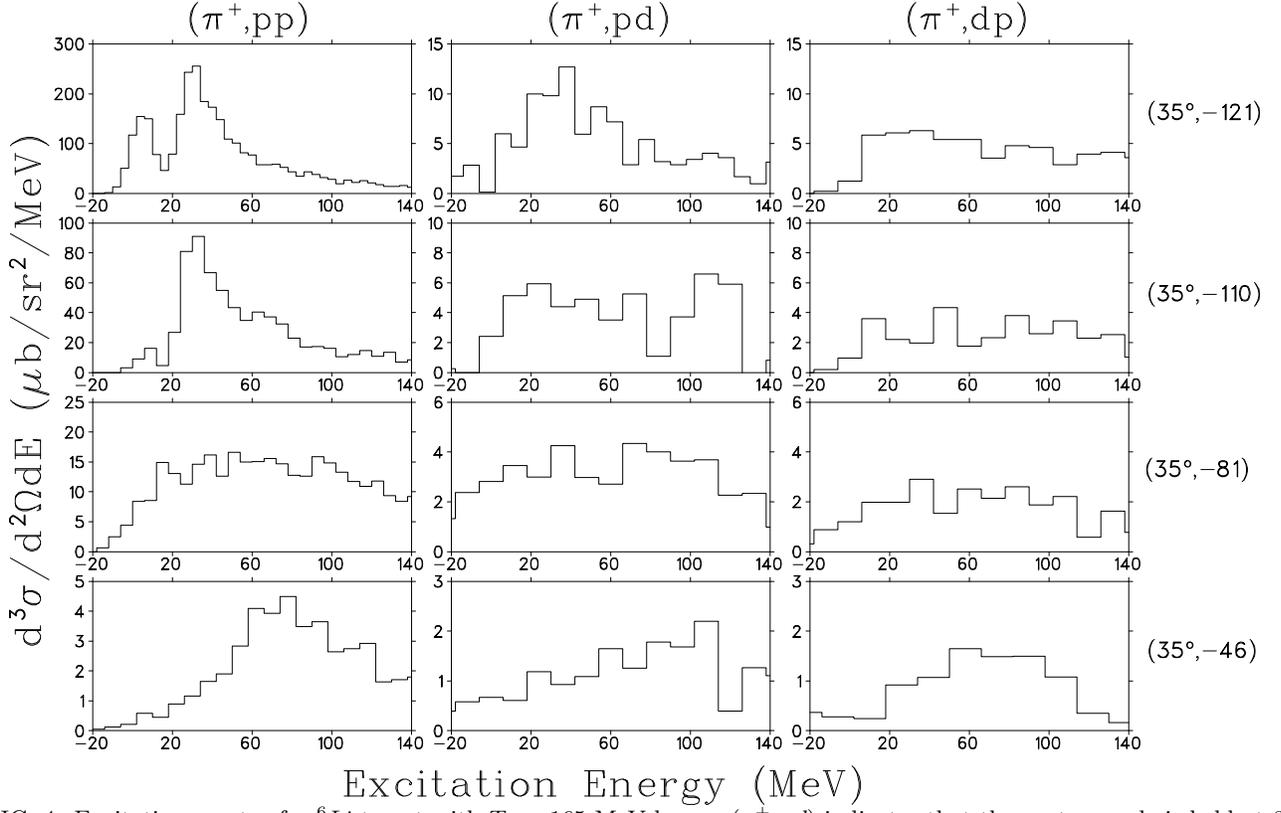,width=4.2in,angle=90}
\caption{Excitation spectra for $^6Li$ target with $T_{\pi}=165$ MeV beam.
  $(\pi^+,pd)$ indicates that the proton angle is held at $35^o$ while the
  deuteron angle is scanned on the other side of the beam, while $(\pi^+,dp)$
  indicates the reverse.  From top to bottom, the rows of histograms correspond
  to $\Delta\theta_{QFA}=4^o$, $15^o$, $44^o$ and $79^o$, respectively.}
\end{center}
\label{fig4}
\end{figure}

To discern the contribution of neutron pickup to the observed deuteron yield,
we first examine the excitation spectra.  In our earlier publications
\cite{lolos96,huber97}, we noted that the single most important determinant of
the $(\pi^+,pp)$ excitation spectra of $^6Li$ and $^{12}C$ was the difference
in angle setting between the two arms and that of $\pi^+d\rightarrow pp$ kinematics,
$\Delta\theta_{QFA}$.  Close to quasi-free absorption kinematics,
$\Delta\theta_{QFA}\approx 0$, the $pp$ excitation spectra exhibited bound
state peaks at low excitation due to the direct emission of protons following
pion absorption.  At $\Delta\theta_{QFA}\ge 30^o$, evidence of more complex
mechanisms (greater than
two-nucleon absorption and multi-step processes), were more apparent.  These
features are repeated in the left columns of figures 4 and 5.

\begin{figure}[h]
\begin{center}
\epsfig{file=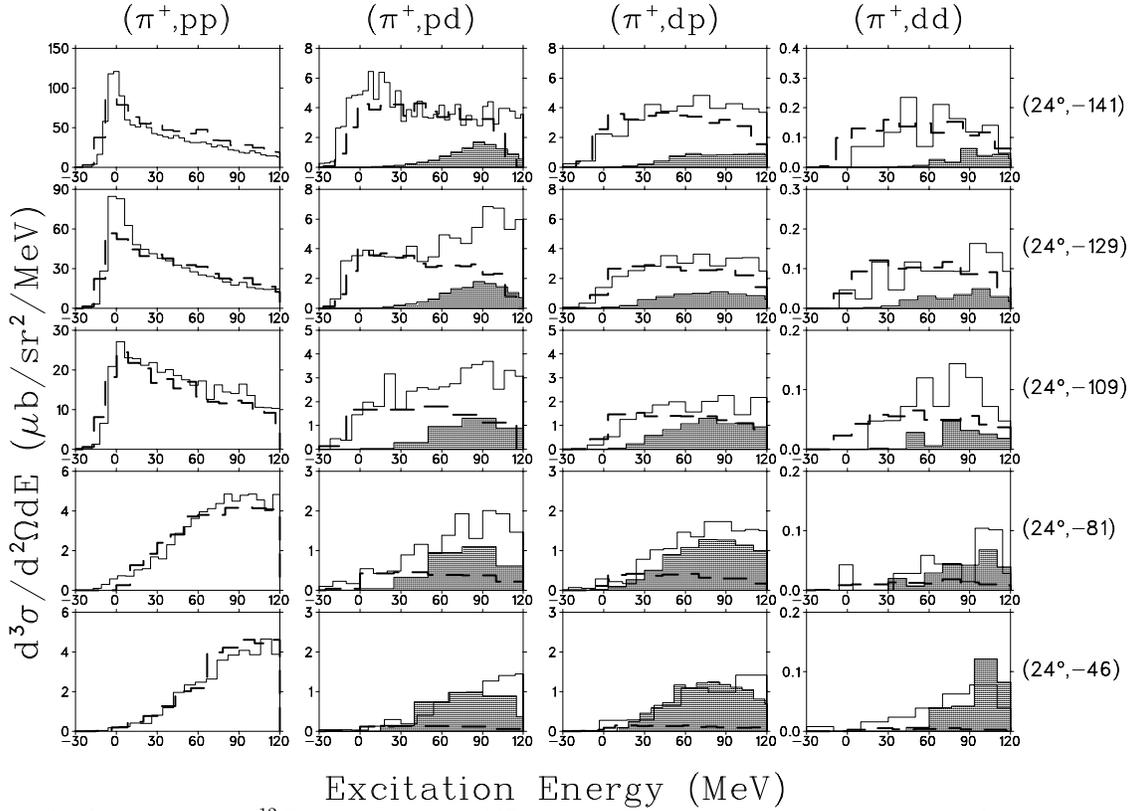,width=4.2in,angle=90}
\caption{Excitation spectra for $^{12}C$ target at $T_{\pi}=165$ MeV.
  From top to bottom, the rows of histograms correspond to
  $\Delta\theta_{QFA}=1^o$, $13^o$, $33^o$, $61^o$, and $96^o$, respectively.
  The solid lines are the experimental data, while the dashed lines on the left
  column are the parameterization of the $pp$ excitation spectra used as input
  to the $pd$ and $dd$ models.  On the three columns on the right, the dashed
  lines indicate the contributions of the $pp$-like model, and the shaded
  regions are those of the pickup model.  All curves associated with a given
  model incorporate the same 165 MeV normalization.}
\end{center}
\label{fig5}
\end{figure}

The dependence of the $(\pi^+,pd)$ excitation spectra on $\Delta\theta_{QFA}$
is much weaker than for $(\pi^+,pp)$.  With the aid of the two models, this weaker dependence can now
be understood.  Figure 5 indicates that the contribution of the $pp$-like model
falls rapidly as one scans away from quasi-free kinematics (i.e. moves from the
top to the bottom of the figure).  However, the contribution of the neutron
pickup mechanism remains more constant with angle, and accounts for essentially
all of the deuteron yield at $\theta_1=24^o$, $\theta_2=-46^o$.  The sum of the
two models provides an excellent description of the data.  These conclusions
also apply to the $(\pi^+,dd)$ data.  It should be noted that the modeled
pickup reaction is highly non-quasi-free in nature, and so contributes to the
$E_{exc}>50$ MeV region, which corresponds to lower emitted deuteron energy.
The contribution of `soft' neutron pickup may be included in the low excitation
region covered by the $pp$-like model.

\begin{figure}[h]
\begin{center}
\epsfig{file=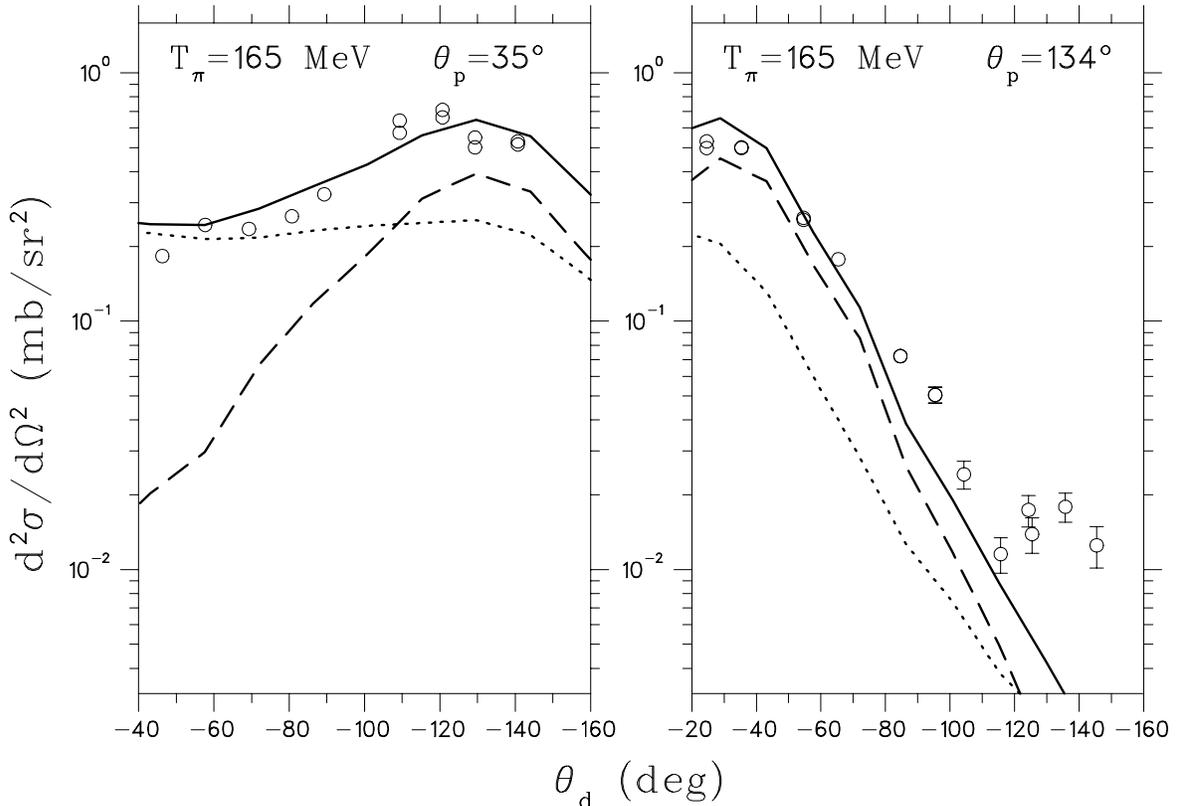,width=4.2in,angle=90}
\caption{Angular distributions of the $^{12}C(\pi^+,pd)$ data at 165 MeV,
  for the proton angle held constant at the value indicated, and the
  deuteron angle scanned on the opposite side of the beam.  The dashed
  curves are the contributions of the $pp$-like model, the dotted curves the
  contributions of the pickup model, and the solid curve the sum of the two. 
  The same normalization factors are used for both panels, and are the same as
  for the corresponding models in figure 5.}
\end{center}
\label{fig6}
\end{figure}

The role of the neutron pickup mechanism is perhaps shown more clearly in
figure 6.  Because the $pn\rightarrow d$ process is strongly forward peaked, it
leads to a significant difference between the angular distributions of forward
angle and backward angle deuteron emission.  In both cases, the observed
deuteron yield due to neutron pickup is the convolution of two factors: the
distribution characteristic of the originally exiting protons, given
approximately by the dashed curve in both panels, and the $pn\rightarrow d$
angular distribution, which is forward peaked in the rest frame of the struck
neutron.  For the leftmost panel, these two factors lead to an almost constant
pickup contribution with deuteron angle, and a very flat $(\pi^+,pd)$ angular
distribution.  This is consistent with the contributions shown in figure 5.  In
the right panel of figure 6, however, the two factors work in concert, and the
neutron pickup contribution falls nearly as fast with angle as the $pp$-like
emission model does.  The contrast between the two experimental angular
distributions shown here is larger than for the equivalent $pp$ emission
distributions shown in reference \cite{huber97}.

In summary, a significant neutron pickup contribution is observed in the
$^{12}C$ data at 165 MeV, accounting for nearly all of the deuteron yield when
both proton and deuteron are detected at forward angles.  This contribution
corresponds to relatively large $^9B^*$ excitation.  For all other angle
combinations, the pickup contribution is smaller than that of the $pp$-like
emission model.  This more detailed description is consistent with the earlier
findings of reference \cite{yokota89}.

\subsection{$T_{\pi}=100$ MeV data}

\begin{figure}[h]
\begin{center}
\epsfig{file=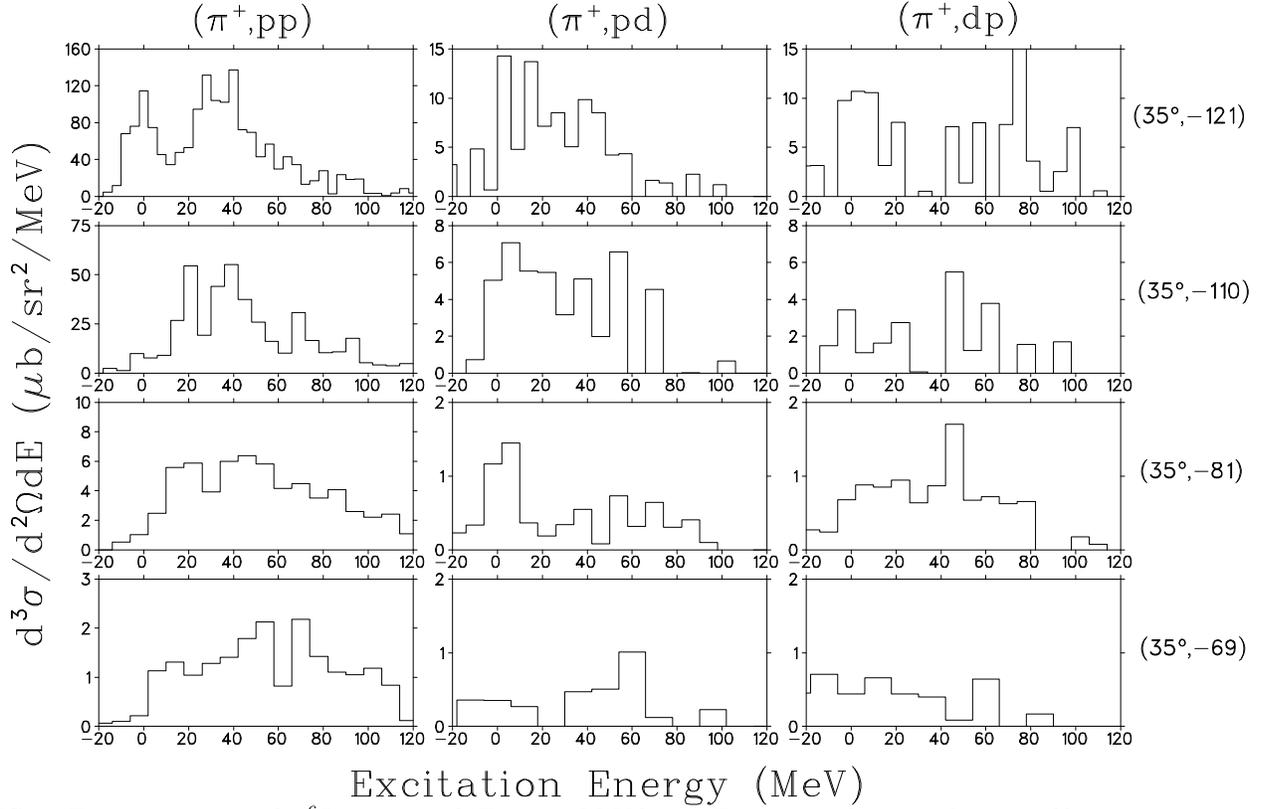,width=4.2in,angle=90}
\caption{Excitation spectra for $^6Li$ target with $T_{\pi}=100$ MeV beam.
From top to bottom, the rows of histograms correspond to
$\Delta\theta_{QFA}=8^o$, $19^o$, $48^o$ and $60^o$, respectively.}
\end{center}
\label{fig7}
\end{figure}

As noted in the introduction, the presence of a pronounced ground state peak in
the $^6Li(\pi^+,pd)$ spectrum at $T_{\pi}=59.4$ MeV \cite{wharton} was
interpreted as evidence of quasi-triton absorption.  Examination of the right
two columns of figures 4, 7 shows that the zero excitation region is slightly
more enhanced at $T_{\pi}=100$ MeV than 165 MeV, possibly indicating a small
quasi-triton contribution to the reaction on $^6Li$ at 100 MeV.  Integrating 
the center column data of figure 7 over all $\theta_d$, the -20 to +20 MeV
excitation region accounts for $37\pm 7\%$ of the observed differential cross
section, while the integral for the center column of the 165 MeV results
(figure 4) accounts for only
$14\pm 2\%$.  A similar enhancement exists for $^{12}C$, with the
equivalent fractions being $36\pm 7\%$, and $15\pm 3\%$ at 100 MeV and 165 MeV,
respectively.

\begin{figure}[h]
\begin{center}
\epsfig{file=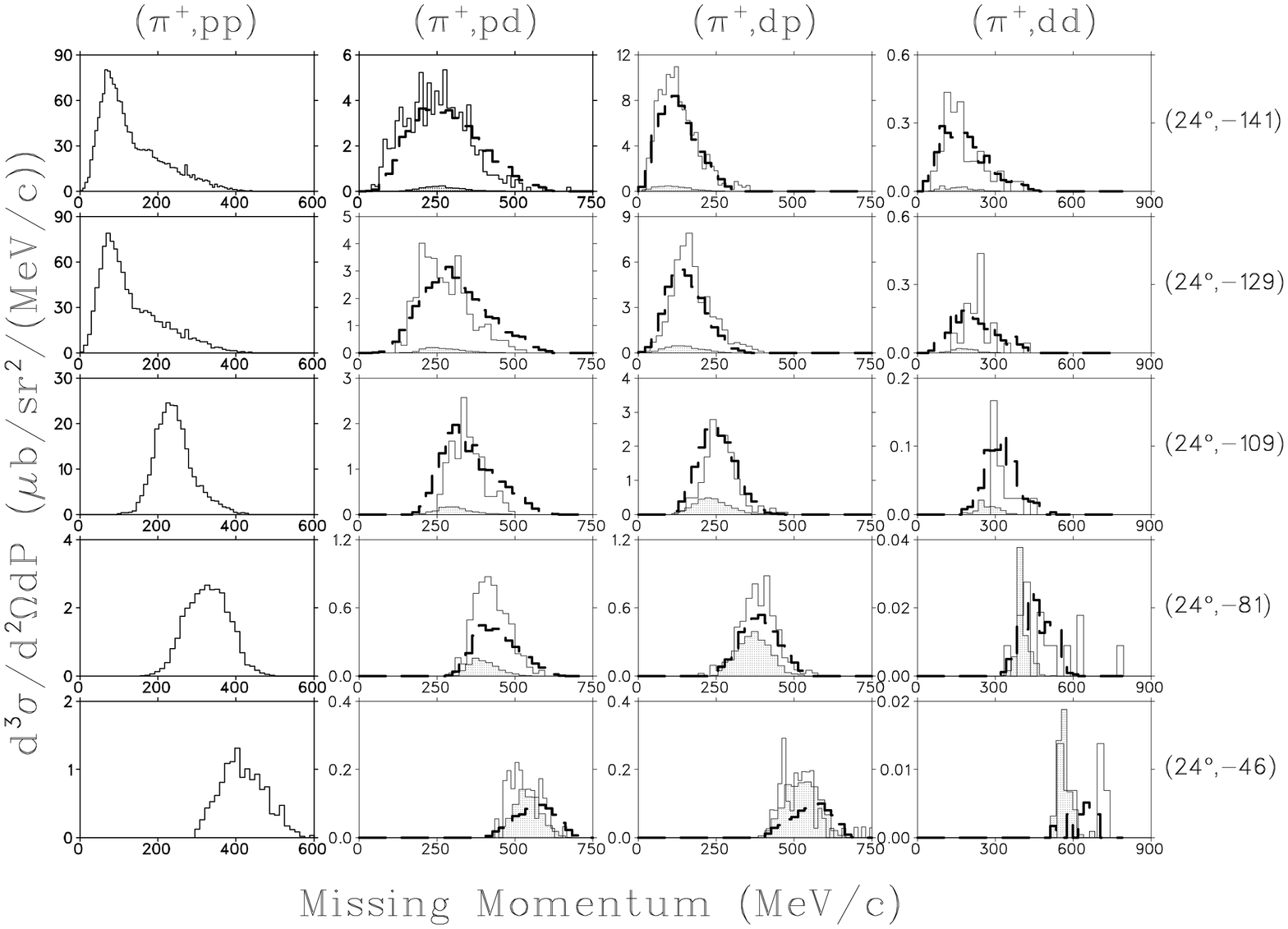,width=4.2in,angle=90}
\caption{Missing momenta spectra for $T_{\pi}=100$ MeV and $^{12}C$ target,
  with $\theta_1$ held at $24^o$ and the second arm scanned as noted.
  The model line types are as in figure 5, and all curves associated with a
  given model incorporate the same 100 MeV normalization.  From top to bottom,
  the rows of histograms correspond to $\Delta\theta_{QFA}=4^o$, $17^o$,
  $37^o$, $65^o$, and $100^o$, respectively.}
\end{center}
\label{fig8}
\end{figure}

A comparison of the $^{12}C$ models and data for the missing momentum spectra
is shown in figure 8.  The histograms for $d$ emission final
states have consistently higher average missing momenta than the $(\pi^+,pp)$
data.  However, since excitation energies of even several hundred MeV are small
compared to the mass of the residual nuclear system, and thus influence the
momentum of the recoiling system only slightly, the distributions of the two
reaction mechanisms overlap substantially for all angle pairs, making the
variable a relatively insensitive indicator.  Nonetheless, the sum of the
models does an excellent job of describing the $pd$, $dp$ and $dd$ emission
data for all angle pairs, with the same, consistent normalization applied.
Generally, the behavior is similar to that seen at 165 MeV, i.e. dominance of
$d$ knockout close to the quasi-free angle, and dominance of $p+n\rightarrow d$
pickup when both arms are at forward angles.

\begin{figure}[h]
\begin{center}
\epsfig{file=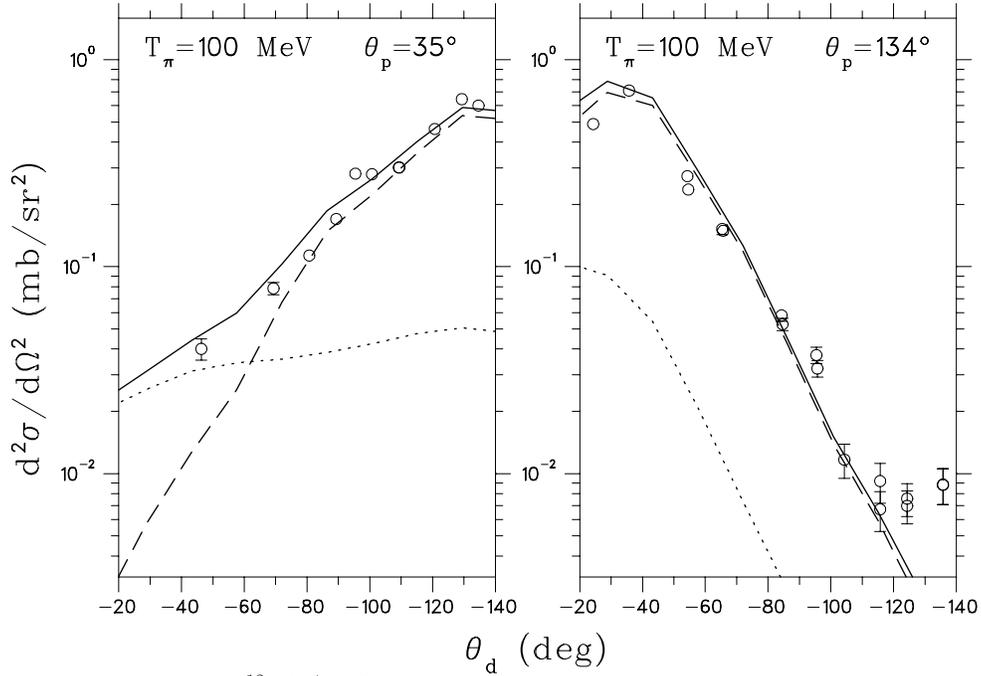,width=3.5in,angle=90}
\caption{Angular distributions of the $^{12}C(\pi^+,pd)$ data at 100 MeV.  The
curves are as in figure 6.  The same normalization factors are used for both
panels, and are the same as for the corresponding models in figure 8.}
\end{center}
\label{fig9}
\end{figure}

\begin{figure}[h]
\begin{center}
\epsfig{file=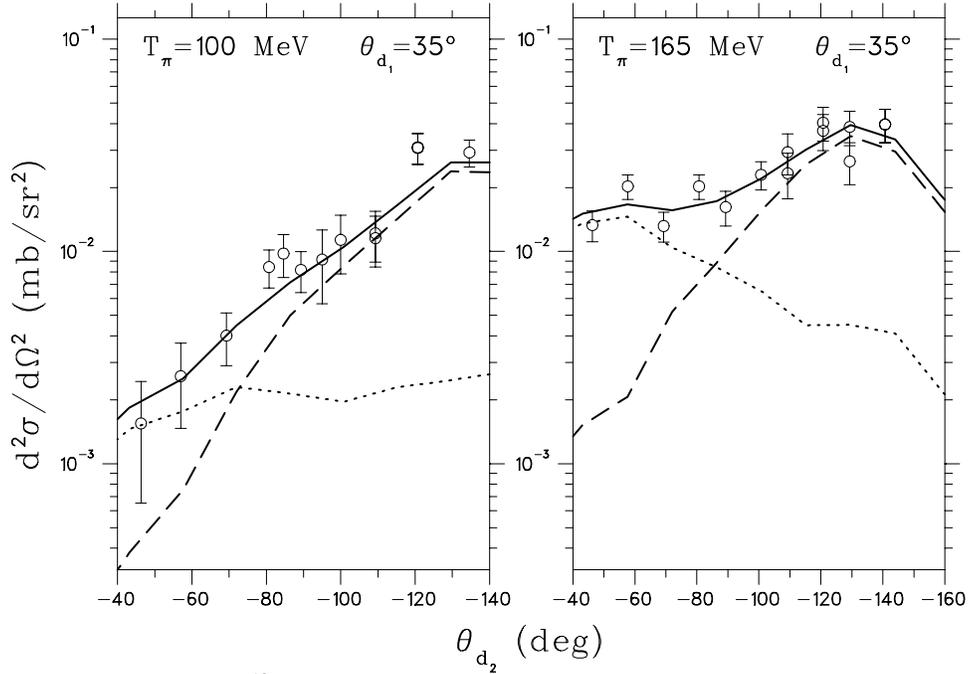,width=3.5in,angle=90}
\caption{Angular distributions of the $^{12}C(\pi^+,dd)$ data.  The curves are
  as in figure 6, with the same normalization factors as used in the right
  columns of figures 8, and 5.}
\end{center}
\label{fig10}
\end{figure}

The difference in the role of the two reaction mechanisms is shown more clearly
in figures 9, 10.  Comparing with figure 6, as the second detector is scanned
away from the quasi-free angle, the angular distributions fall more rapidly at
100 MeV than at 165 MeV.  This is due to the smaller role of the pickup
mechanism at 100 MeV, a finding consistent with domination of the
quasi-triton mechanism at 59.5 MeV incident pion energy \cite{wharton}.
However, because the cross section of the $^{12}C(p,d)^{11}C_{g.s.}$ reaction
falls with energy, from 16 mb at $T_p=30.3$ MeV, to 1.5 mb at $T_p=156$ MeV
\cite{pd30,pd156}, the argument can be made that the neutron pickup
contribution should also fall with increasing incident pion energy.  The
observed, opposite behavior is almost certainly due to the greater multiplicity
of the reaction at the higher pion energy \cite{rowntree}.  It should be noted
that the effect of the significant experimental detection threshold does not
change this result in any substantive manner.  Since the neutron pickup events
are characterized by a relatively high excitation energy, and hence lower
deuteron energy, these events suffer from the experimental threshold more
severely, with a detection acceptance less than half that of the $pp$-like
events.  While both mechanisms have higher acceptances at 165 MeV than at 100
MeV, the $n$ pickup acceptance is lower by only an additional 17\% relative to
the $pp$-like events.

In summary, the experiment indicates a larger direct quasi-triton absorption
component to the reaction at 100 MeV compared to 165 MeV.  Conversely, the
contribution of the hard neutron pickup process is smaller at this energy.

\subsection{Single differential and total cross sections}

The double differential cross sections were fitted with Legendre polynomials
and integrated in order to obtain $d\sigma/d\Omega$.  Checks were made
to ensure that the result was not sensitive to the highest order employed in the
fits.  Figure 11 displays the resulting single
differential cross section angular distributions.

At both 100 and 165 MeV, the forward angle differential cross sections on
$^6Li$ exceed those on $^{12}C$, and the angular distribution has a large
$30^o/90^o$ cross section ratio, resembling that of the $\pi^+d\rightarrow pp$
reaction.  The figure also shows the estimated neutron pickup contribution for
$^{12}C$, plotted versus proton emission angle.  The angular distribution of
the neutron pickup contribution is especially forward peaked in the $^{12}C$
165 MeV panel, and if it were subtracted from the total $^{12}C(\pi^+,pd)$
distribution (solid curve), the result would be an angular distribution more
similar in shape to the $^6Li$ distribution.  The neutron pickup mechanism
contributes approximately 50\% of the $pd$ total cross section on $^{12}C$ at
165 MeV, but only 20\% at $T_{\pi}=100$ MeV.

\begin{figure}[h]
\begin{center}
\epsfig{file=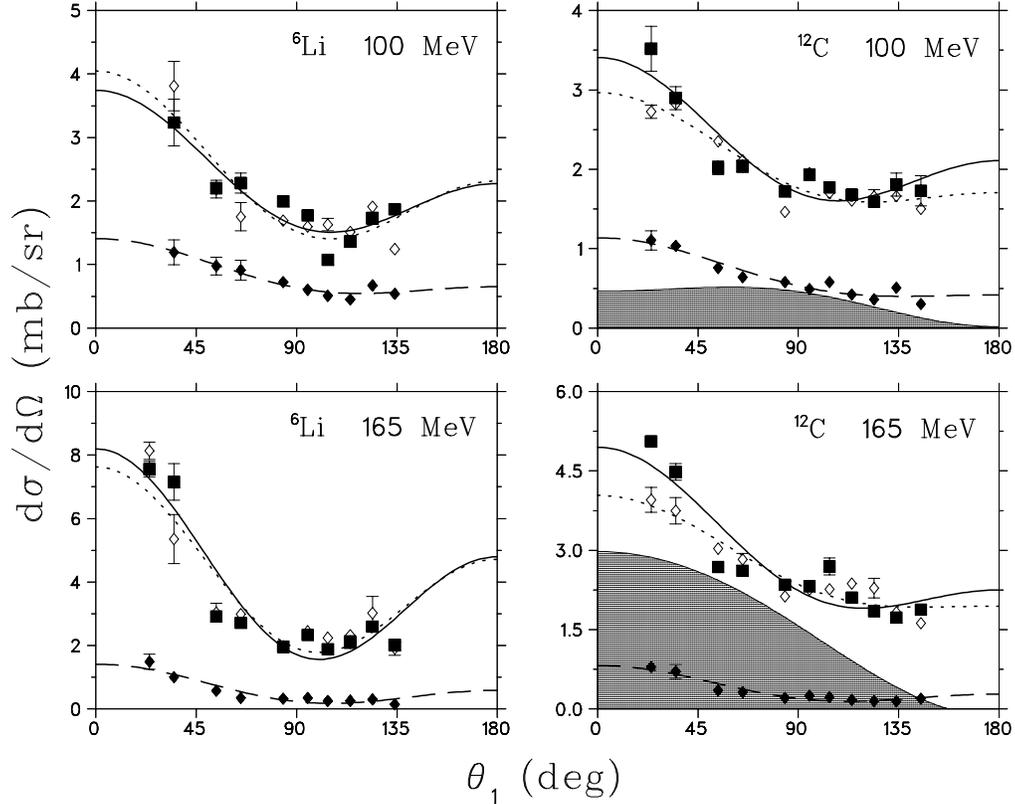,width=4.2in,angle=90}
\caption{Lab frame differential cross sections for the $(\pi^+,pd)$
[solid squares] and $(\pi^+,dp)$ [open diamonds] reactions on $^6Li$
and $^{12}C$.  The solid [dotted] line is a second order Legendre
polynomial fit to the $(\pi^+,pd)$ [~$(\pi^+,dp)$~] angular distributions
while the shaded regions are the estimated neutron pickup contributions, all plotted
versus $\theta_p$ and to be compared against the solid curves.  Finally, the
solid diamonds are $(\pi^+,pd)$ differential cross sections with the
final state excitation restricted to $0\pm 20$ MeV.  These cross sections do
not include contributions below the detection threshold of the experiment.}
\end{center}
\label{fig11}
\end{figure}

Direct quasi-triton absorption should contribute primarily to the +20 to -20
MeV excitation region, and so data restricted to this region are also shown
in the figure.  The Legendre polynomial integrated results over all angles
indicate that this excitation region accounts for $37\pm 6\%$ and $14\pm 4\%$
of the total $pd$ cross section on $^6Li$ at 100 and 165 MeV, respectively, and
$30\pm 4\%$, $12\pm 3\%$ on $^{12}C$.  Using $T_{\pi}=65$ MeV pions on
$^{16}O$ target, Bauer et al. \cite{bauer,hamers} reported a significantly
different angular distribution for the $(\pi^+,pd)$ reaction than the
$(\pi^+,pp)$ reaction.  While the $pp$ events exhibited the typical
$cos^2\theta$ dependence known for quasi-deuteron absorption, the $pd$ events
had a highly asymmetric distribution, with forward-going protons and
backward-going deuterons being the most likely combination.  This was explained
in terms of direct $\pi^+t\rightarrow pd$ absorption.  In our case, the dotted
and dashed curves of each panel are very similar, indicating that the proton
and deuteron emission distributions are essentially identical.  This reinforces
the conclusion that the quasi-triton absorption component of the reaction is
relatively small.

\begin{figure}[h]
\begin{center}
\epsfig{file=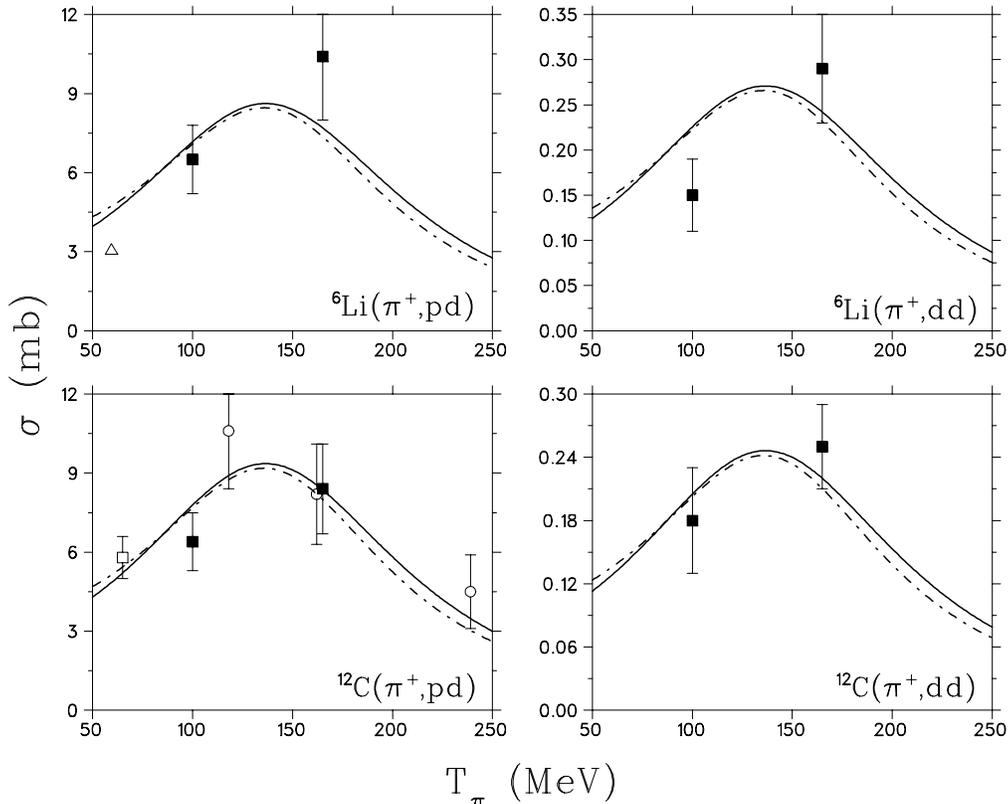,width=4.2in,angle=90}
\caption{Total cross sections for this work (solid squares),
to discrete states of $^3He$ only \protect\cite{wharton} (triangle), on
$^{16}O$ \protect\cite{bauer} (empty square), and on $^{14}N$
\protect\cite{rowntree} (circle), the latter two scaled by $A^{2/3}$ to
$^{12}C$ equivalent.  The error bars for this work include all statistical and
systematic uncertainties.  In the case of the Ref. \protect\cite{rowntree}
results, the value shown is the sum of their ``2 charged multiplicity'' cross
section for 1 deuteron and any number of neutrons detected, since neutrons are
undetected by our apparatus.  The solid curve is the parameterization of the
$\pi^+d\rightarrow pp$ reaction provided by Ritchie \protect\cite{ritchie2},
while the dash-dot curve is the parameterization by VerWest and Arndt
\protect\cite{verwest}.  From upper left to lower right panel, the curves are
normalized by factors of 0.70, 0.022, 0.76, and 0.020, respectively.}
\end{center}
\label{fig12}
\end{figure}

$(\pi^+,pd)$ and $(\pi^+,dd)$ total cross sections were obtained by fitting the
single differential cross sections with the Legendre polynomial distributions
shown in figure 11, to yield the values shown in figure 12.  This is the same
method as used in our earlier publications \cite{lolos96,huber97} and does not
include corrections for final state interactions.  It is evident that the $pd$ total
cross sections have an energy dependence that is broadly similar to the
$\pi^+d\rightarrow pp$ reaction.

\section{Summary and Conclusions}

In this experiment, $(\pi^+,pd)$, and $(\pi^+,dd)$ reactions were investigated
using 100 and 165 MeV pion beams incident on $^{12}C$ and $^6Li$ targets,
bringing to a close our investigation of the pion absorption reaction with
discrete detectors at TRIUMF.  It was observed that the $pd$ angular
distributions were flatter than the $pp$ distributions, and that the $dd$ were
the flattest of all.  The two sets of distributions were offset from each other
by nearly the same factor, indicating that the deuteron emission mechanism has
the same origin in the two cases.

The $pd$, $dp$ and $dd$ kinematic distributions were compared with two
empirical models.  The first model, based on a parameterization of the full
excitation spectrum of our earlier published \cite{huber97} $^{12}C(\pi^+,pp)$
data, was termed `pp-like' emission, as it assumes that the deuteron emission
energy and angle distributions mirror those of proton emission.  This could
either be due to direct deuteron emission following $\pi^+$ absorption,
deuteron knockout via ISI, or neutron pickup following a soft FSI.  The second
model, which included $^{12}C(p,d)^{11}C_{g.s.}$ cross sections as one of its
inputs, simulated neutron pickup following $pp$ emission via a hard FSI.  The
two mechanisms were fit to the data, with a common normalization
factor for each mechanism at each of the two incident pion energies studied,
independent of proton or deuteron emission angle.  It was found that the
appropriately normalized sum of the two mechanisms were able to describe all of
the features of the data in an acceptable manner, and that nearly all of the
observed yield for forward protons in coincidence with forward deuterons was
due to neutron pickup.  The importance of this mechanism increased with energy,
possibly due to the increasing multiplicity of the underlying $\pi^+$
absorption reaction.  Signatures of deuteron emission following direct
quasi-triton absorption were also investigated, and were only found to play a
substantive role at $T_{\pi}=100$ MeV.

It is now known that the mechanisms contributing to pion absorption are even
more complex than originally anticipated \cite{lehmann,huber97}, so the clean
identification of the role of FSI in such reactions remains a
difficult task.  The kinematic signatures of hard FSI are not distinctive, and
occupy regions of phase-space already populated by other complex absorption
mechanisms.  Fortunately, however, neutron pickup has a distinctive
experimental signature, and it is highly likely that it is one of the many
possible hard final-state interactions following absorption.  This work
reaffirms the significant role played by FSI in hadronic interactions, and our
success in quantitatively reproducing the shapes of the experimental
distributions in terms of the $p-n$ pickup mechanism leads one to
optimistically expect that the role of FSI in pion absorption reactions will
ultimately be understood in much better detail than at present.  FSI, of
course, are not only important in hadron-induced reactions, but also in
electron and photon-induced reactions.  The issue of modeling other FSI, such
as $NN$ rescattering in all such reactions, could be revisited using a
semi-empirical approach similar to that used here.

\section{Acknowledgements} 

We would like to thank S. Rusaw and M. Jackson for assisting in
the analysis.  This project was supported in part by the Natural Sciences and
Engineering Research Council of Canada (NSERC), and the Saskatchewan Department
of Economic Development. The generous assistance of E.W. Vogt, then Director of
TRIUMF, is also gratefully acknowledged.



\begin{references}
\bibitem[(a)]{greece} Present Address: Department of Physics, University of
Ioannina, GR-45110 Ioannina, Greece
\bibitem{ashery} D. Ashery, J.P. Schiffer, Ann. Rev. Nucl. Part. Sci. {\bf 36}
  (1986) 207.
\bibitem{lads} D. Androic, et al., Phys. Rev. C {\bf 53} (1996) R2591.\\
G. Backenstoss, et al., Phys. Lett. {\bf B 379} (1996) 60.
\bibitem{comiso} J.C. Comiso, et al., Phys. Lett. {\bf 78B} (1978) 31.
\bibitem{wharton} W.R. Wharton, et al., Phys. Rev. C {\bf 33} (1986) 1435.
\bibitem{yokota86} H. Yokota, et al., Phys. Lett. B {\bf 175} (1986) 23.
\bibitem{yokota89} H. Yokota, et al., Phys. Rev. C {\bf 39} (1989) 2090.
\bibitem{ransome90} R.D. Ransome, et al., Phys. Rev. C {\bf 42} (1990) 1500.
\bibitem{ransome92} R.D. Ransome, et al., Phys. Rev. C {\bf 45} (1992) R509.
\bibitem{bauer} Th.S. Bauer, et al., Phys. Rev. C {\bf 46} (1992) R20.
\bibitem{rowntree} D. Rowntree, et al., Phys. Rev. C {\bf 60} (1999) 054610.
\bibitem{lehmann} A. Lehmann, et al., Phys. Rev. C {\bf 55} (1997) 2931.
\bibitem{pap95} Z. Papandreou, et al., Phys. Rev. C {\bf 51} (1995) R2862.
\bibitem{lolos96} G.J. Lolos, et al., Phys. Rev. C {\bf 54} (1996) 211.
\bibitem{huber97} G.M. Huber, et al., Nucl. Phys. {\bf A 624} (1997) 623.
\bibitem{pap88} Z. Papandreou, G.J. Lolos, G.M. Huber and X. Aslanoglou, Nucl.
Instr. Meth. {\bf A268} (1988) 179.
\bibitem{pap88a} Z. Papandreou et. al., Nucl. Instr. Meth. {\bf B34} (1988)
454. 
\bibitem{gooding} T.J. Gooding and H.G. Pugh, Nucl. Instr. Meth. {\bf 7} (1960) 
189.
\bibitem{pd30} N.S.Chant, P.S.Fisher, D.K.Scott, Nucl. Phys. {\bf A99}
  (1967) 669.
\bibitem{pd52} H.Ohnuma et al., J. Phys. Soc. Jpn. {\bf 48} (1980)
  1812.
\bibitem{pd65} P.G.Roos, et al., Nucl. Phys. {\bf A255} (1975) 187.
\bibitem{genbod} CERN Computer Library W 505, 1977.
\bibitem{pd100} J.K.P.Lee, et al., Nucl. Phys. {\bf A106} (1968) 357. 
\bibitem{pd156} D. Bachelier, et al., Nucl. Phys. {\bf A126} (1969) 60.
\bibitem{pd185} J. Kallne, E. Hagberg, Phys. Scr. {\bf 4} (1971) 151.
\bibitem{hamers} R. Hamers, Ph.D. thesis, Free University of
  Amsterdam, 1988, unpublished.
\bibitem{ritchie2} B.G. Ritchie, Phys. Rev. C {\bf 44} (1991) 533.
\bibitem{verwest} B.J. VerWest and R.A. Arndt, Phys. Rev. C {\bf 25}
(1982) 1979.
\end{references}
\end{document}